\documentstyle[jkas,graphicx]{article}

\beginpage{}
\endpage{}
\year{2008} \volume{}

\runningauthor {Y. TAKEDA, ET AL.} \year{2008} \volume{}
\beginpage{1}\endpage{}
\runningtitle{ROTATION AND SURFACE ABUNDANCE PECULIARITIES 
IN A-TYPE STARS}

\begin{document}
\title{ROTATION AND SURFACE ABUNDANCE PECULIARITIES \\
IN A-TYPE STARS}
\author{Yoichi Takeda$^1$, Inwoo Han$^2$, Dong-Il Kang$^3$,
Byeong-Cheol Lee$^{2,4}$, and Kang-Min Kim$^2$}
\address{$^1$ National Astronomical Observatory of Japan,
2-21-1 Osawa, Mitaka, Tokyo 181-8588, Japan\\
 {\it e-mail}: takeda.yoichi@nao.ac.jp}
\address{$^2$ Korea Astronomy and Space Science Institute,
61-1 Whaam-dong, Youseong-gu, Taejon 305-348, Korea\\
 {\it e-mail}: iwhan@kasi.re.kr, bclee@boao.re.kr, kmkim@boao.re.kr}
\address{$^3$ Gyeongsangnamdo Institute of Science Education,\\
75-18 Gajinri, Jinsungmyeon, Jinju, Gyeongnam 660-851, Korea \\
 {\it e-mail}:  kangdongil@gmail.com}
\address{$^4$ Department of Astronomy and Atmospheric Sciences,\\ 
Kyungpook National University, Daegu 702-701, Korea \\}

\address{\normalsize{\it (Received  ;   Accepted)}}
\offprints{Y. Takeda}
\abstract{
In an attempt of clarifying the connection between the
photospheric abundance anomalies and the stellar rotation
as well as of exploring the nature of ``normal A'' stars,
the abundances of seven elements (C, O, Si, Ca, Ti, Fe, and Ba) 
and the projected rotational velocity for 46 A-type field stars 
were determined by applying the spectrum-fitting method to 
the high-dispersion spectral data obtained with BOES at BOAO.
We found that the peculiarities (underabundances of 
C, O, and Ca; an overabundance of Ba) seen in slow rotators 
efficiently decrease with an increase of rotation, which almost 
disappear at $v_{\rm e}\sin i \ga 100$~km~s$^{-1}$. 
This further suggests that stars with sufficiently large rotational
velocity may retain the original composition at the surface without
being altered.
Considering the subsolar tendency (by several tenths dex below) 
exhibited by the elemental abundances of such rapidly-rotating 
(supposedly normal) A stars, we suspect that 
the gas metallicity may have decreased since our Sun was born, 
contrary to the common picture of galactic chemical evolution.
}


\maketitle

\section{INTRODUCTION}

Since unevolved A-type stars on (or near to) the upper main-sequence  
have masses around $\sim 2 M_{\odot}$, their surface abundances 
may retain information of the composition of the past galactic gas 
($\la 10^{9}$~yr ago) from which they formed; this would provide us
with an important opportunity to investigate the ``recent''
chemical evolution history of the Galaxy.
However, such a study using A stars as a probe of late-time history
of chemical evolution has rarely been done in spite of its potential 
significance\footnote{For example, Takeda, Sato, \& Murata (2008) 
found in their extensive study of late-G giants
(also having masses around $\sim 2 M_{\odot}$; evolved counterparts
of A dwarfs) that their metallicities show an appreciable diversity
as large as $\sim 1$~dex ($-0.8 \la$~[Fe/H]~$\la +0.2$) with a
subsolar trend on the average (cf. Fig. 14 therein), from which
they argued that some special event (like a mixing of metal-poor
primordial gas caused by infall) might have occurred $\sim 10^{9}$~yr
ago. However, before making any speculation, it is important to 
confirm whether such a trend is also observed in their progenitors 
in the upper main-sequence.}
in contrast to the case of old-time history where a number of 
researches (using longer-lived F--G--K dwarfs) are available.
This is related to the fact that a large fraction of them are rapid 
rotators (typically $v_{\rm e} \sin i \sim$ 100--200~km~s$^{-1}$ 
on the average; see, e.g., Royer et al. 2002a, b) 
whose spectra are technically difficult to analyze 
because lines are broad and smeared out, while sharp-lined slow 
rotators easy to handle tend to show abundance anomalies (chemically
peculiar stars or CP stars).\footnote{Although many things are 
left unresolved concerning the origin and nature of the CP phenomena, 
it is widely considered (at least in the qualitative sense) that 
the chemical segregation in the stable atmosphere is responsible 
for the abundance anomalies at the surface: i.e., an element becomes 
over- or under-abundant depending on the balance of upward radiation 
force and the downward gravitational force. In this case, rotation 
would act against an efficient built-up of such anomalies because 
it enhances mixing of outer stellar layers via shear instability or 
meridional circulation.}
As a matter of fact, most spectroscopic analyses of A stars  
have focused on sharp-lined ones with $v_{\rm e}\sin i \la 50$~km~s$^{-1}$.
For this reason, nobody could be sure whether the result obtained 
for star classified as ``normal A'' really reflects the initial 
composition free from any peculiarities. 

Therefore, in order to make a further step forward, 
it is requisite to challenge abundance determinations for
``unbiased'' sample of A-type stars in general 
(i.e., without sidestepping rapidly rotating ones), 
which inevitably requires an application of
the spectrum synthesis technique, since reliably measuring
the equivalent widths of individual spectral lines is 
almost hopeless for rapid rotators. 
Admittedly, while such trials of determining abundances from 
spectra of A dwarfs including broad-lined ones have recently emerged 
thanks to the improvement in the method of analysis as well as 
the data quality, their interests are mainly directed to objects 
of specific types; e.g., Vega-like stars (Dunkin et al. 1997) or 
$\lambda$ Bootis stars (Andrievsky et al. 2002) or 
open-cluster stars (Takeda \& Sadakane 1997; Varenne \& Monier 1999; 
Gebran et al. 2008; Gebran \& Monier 2008; Fossati et al. 2008). 
Namely, a systematic study attempting to clarify the characteristics 
of normal field A-type stars in general, especially in terms of their 
abundance--rotation connection, seems to have been rarely attempted. 
To our knowledge, only one such study is Lemke's (1990, 1993)
determinations of C and Ba abundances for some 20 rapidly-rotating 
A stars with $v_{\rm e} \sin i$ up to $\sim$~200~km~s$^{-1}$, 
which however appear to be still insufficient and inconclusive 
as judged from his adopted method of approach as well as 
the number of elements studied. 

Considering this situation, we decided to revisit this problem
in our own manner based on the high-dispersion spectral data of 
$\sim 50$ A-type stars in a wide range of $v_{\rm e} \sin i$ 
(0--300~km~s$^{-1}$) obtained with BOES at BOAO, while applying 
the automatic spectrum fitting algorithm (Takeda 1995) which 
efficiently enables determinations of the abundances
(for selected six elements of C, O, Si, Ti, Fe, and Ba)
even for rapid rotators showing considerably merged spectra. 
Our ultimate aim is to clarify the following questions of interest:\\
--- (1) Is there any systematic rotation-dependent tendency 
between slow and rapid rotators in terms of the abundance anomaly? 
If so, what is the critical value of $v_{\rm e} \sin i$, above which
 stars may be regarded as normal? \\
--- (2) What would the abundance characteristics of
``normal A-type stars'' like, which we may consider as
retaining the composition of the galactic gas from which
they formed? \\

We will show that reasonable answers to these points are provided 
from this study (cf. Sect. V).

\section{OBSERVATIONAL DATA}

We selected 46 apparently bright ($V \la 5$~mag) A-type stars 
(including Am stars\footnote{Although we excluded SrCrEu-type 
Ap stars (many are known to have magnetic fields) 
as well as $\lambda$ Boo stars (dust--gas separation process 
may be responsible for their metal-deficient trend),
which show outstandingly large abundance peculiarities, Am stars 
(metallic-line stars which show comparatively mild anomaly) were 
included in our list (as was done by Takeda \& Sadakane 1997), 
since otherwise we can not realize a statistically meaningful 
sample of stars with a wide range of $v_{\rm e}\sin i$; i.e., the number 
of stars with small $v_{\rm e} \sin i$ which are not classified as Am 
is too small. (Since Am peculiarity is considered to be a natural 
phenomenon accompanied by slow rotation, Am stars would rather be 
interpreted as ``ordinary'' slowly-rotating A stars).}) as our targets,
which are listed in Table 1. Figure 1 shows the plots of these 
stars on the theoretical HR diagram, where we can see that 
their masses are in the range of 
$1.5 M_{\odot} \la M \la 3 M_{\odot}$.

The observations were carried out on 2008 January 14--16
by using BOES (Bohyunsan Observatory Echelle Spectrograph)
attached to the 1.8 m reflector at Bohyunsan Optical 
Astronomy Observatory. 
Using 2k$\times$4k CCD (pixel size of 15~$\mu$m~$\times$~15~$\mu$m), 
this echelle spectrograph enabled us to obtain spectra of wide 
wavelength coverage (from $\sim$~3700~$\rm\AA$ to 
$\sim$~10000~$\rm\AA$) at a time.
We used 200$\mu$m fiber corresponding to the resolving power 
of $R \simeq 45000$. The integrated exposure time 
for each star was typically $\sim$~10--20~min on the average.
 
The reduction of the echelle spectra (bias subtraction, flat 
fielding, spectrum extraction, wavelength calibration, and 
continuum normalization) was carried out with the software 
developed by Kang et al. (2006). For all 46 targets, we could
accomplish S/N ratio of $\sim$~300--500 at the 6150~$\rm\AA$ 
region (on which we placed the largest weight on abundance
determination; cf. Sect. IV). 

\section{ATMOSPHERIC MODELS}

The effective temperature ($T_{\rm eff}$) and the surface gravity
($\log g$) of each program star were determined from the colors
of Str\"{o}mgren's $uvby\beta$  photometric system with the help 
of the {\tt uvbybetanew}\footnote{
Available at http://www.astro.le.ac.uk/\~{}rn38/uvbybeta.html.}
program (Napiwotzki, Sch\"{o}nberner, \& Wenske 1993),
which is an updated/combined version of Moon's (1985) 
{\tt UVBYBETA} (for dereddening) and {\tt TEFFLOGG} (for
determining $T_{\rm eff}$ and $\log g$) codes while based on 
Kurucz's (1993) ATLAS9 models.
The observed color data ($b-y$, $c_{1}$, $m_{1}$, $\beta_{1}$)
of each star were taken from the extensive compilation of 
Hauck \& Mermilliod (1980) via the SIMBAD database. The resulting 
values of $T_{\rm eff}$ and $\log g$ are given in Table 1.

Then, the model atmosphere for each star was constructed
by two-dimensionally interpolating Kurucz's (1993) ATLAS9 
model grid in terms of $T_{\rm eff}$ and $\log g$, where
we exclusively applied the solar-metallicity models 
as was done in Takeda \& Sadakane (1997) or Takeda et al. (1999).

\section{ANALYSIS}

\subsection{Method and Selected Regions}

As a numerical tool for extracting information from the spectra,
we adopted the multi-parameter fitting technique developed by
Takeda (1995), which can simultaneously determine various 
parameters affecting the spectra; e.g., abundances of elements 
showing lines of appreciable contributions, the projected rotational 
velocity, or the microturbulent velocity.

In the present study, we decided to concentrate on three 
wavelength regions to be analyzed:
(1) 6140--6170~$\rm\AA$ region (hereinafter called ``6150 region'')
including lines of O, Si, Ca, Fe, and Ba; (2) 5375--5390~$\rm\AA$ 
region (``5380 region'') including lines of C, Ti, and Fe; 
(3) 7765--7785~$\rm\AA$ region (``7775 region'') including lines
of O and Fe.

\subsection{Microturbulence}

One of the important key parameters is the microturbulence ($\xi$),
the choice of which can be critical in abundance determinations
from strong line features. Although our method of analysis
provides us with a possibility of establishing this parameter
as demonstrated by Takeda (1995), whether it works successful 
or not depends upon situations (i.e., not always possible;
especially, its difficulty grows as the rotation becomes higher).
Besides, we found from experiences that solutions can sometimes
converge at inappropriate (or erroneous) $\xi$ values
for the cases of rapid rotators or insufficient data quality.

Accordingly, on the supposition that $\xi$ is a function of
$T_{\rm eff}$, we decided to find an appropriate analytical
formula by combining the solutions of $\xi$ for the 
successfully determined cases. For this purpose, special 
preparatory multi-parameter fitting analyses ``including $\xi$ 
as a variable'' (in addition to the elemental abundances and 
the rotational velocity focused in the standard analysis described 
in Sect. IV-c) were first carried out for the 6150 region and 
the 7775 region with an intention to derive the $\xi^{\rm fit}$ values 
(denoted as $\xi^{\rm fit}_{6150}$ and $\xi^{\rm fit}_{7775}$, respectively).
It turned out that $\xi^{\rm fit}_{6150}$ and $\xi^{\rm fit}_{7775}$ 
could be determined for 33 and 13 stars, respectively, out of
46 program stars.

The correlation of $\xi^{\rm fit}_{6150}$ and $\xi^{\rm fit}_{7775}$ 
for 13 stars in common is displayed in Figure 2a, where we do not recognize
any systematic discordance between these two, though the scatter
is rather large (the average difference is $\sim 0.1$~km~s$^{-1}$ 
with the standard deviation of $\sim 1$~km~s$^{-1}$).
Figure 2b shows these $\xi^{\rm fit}_{6150}$ and $\xi^{\rm fit}_{7775}$ 
values plotted against $T_{\rm eff}$, where the $\xi$ results derived 
for F--G--K dwarfs/subgiants are also overplotted for comparison.
We can see from this figure that, as  $T_{\rm eff}$ becomes higher,
$\xi^{\rm fit}$ increases from $\sim 1$~km~s$^{-1}$ (at $T_{\rm eff} \sim 6000$~K),
to its nearly maximum value of $\sim 4 (\pm 2) $~km~s$^{-1}$ (at 
$T_{\rm eff} \sim 8000$~K) though with a considerably
large scatter, followed by a decreasing tendency toward higher 
$T_{\rm eff}$ of $\sim 10000$~K (where we know $\xi$ is typically
$\sim$~1--2~km~s$^{-1}$; cf. Sadakane 1990).
Hence, we adopt an analytical formula for the standard microturbulence
($\xi^{\rm std}$)
\begin{equation}
\xi^{\rm std} = 4.0 \exp\{- [\log (T_{\rm eff}/8000)/A]^{2}\} \\
\end{equation}
(where $A \equiv [\log (10000/8000)]/\sqrt{\ln 2}$)
with probable uncertainties of $\pm 30\%$, which roughly represents
(and encompasses) the observed trend as shown in Figure 2b.
Note that such $\xi^{\rm std}$ vs. $T_{\rm eff}$ relation we have defined
is in reasonable agreement with previous results (see, e.g., 
Figure 1 in Coupry \& Burkhart 1992 or Figure 2 in 
Gebran \& Monier 2007).
The $\xi$ values for each of the 46 stars evaluated by Equation (1),
which we will use for abundance determinations, are given in Table 1.

\subsection{Solutions}

Now that the model atmosphere and the microturbulence 
have been assigned to each star, we can go on to
evaluations of elemental abundances by way of synthetic spectrum
fitting applied to three wavelength regions. 
All the atomic data (wavelength, excitation potential, oscillator
strengths, damping constants) relevant to the analysis were 
taken from the extensive compilation of Kurucz \& Bell (1995),
except for $\log gf$(Fe~{\sc i} 7780.552) (cf. the caption of Table 2).
The adopted data of important lines are summarized in Table 2.
The non-LTE effect was explicitly taken into account only for
the O~{\sc i} triplet lines at 7771--5~$\rm\AA$ (for which
the non-LTE correction is known to be appreciably large and
its inclusion is necessary; see, e.g., Takeda \& Sadakane 1997) 
based on the statistical equilibrium calculation for O~{\sc i}
(cf. Takeda 2003); otherwise, we assumed LTE. 

Applying our automatic fitting approach, we adjusted
the following parameters to accomplish the best fit at each region:
$A^{\rm O}_{6150}$, $A^{\rm Si}_{6150}$, $A^{\rm Ca}_{6150}$, 
$A^{\rm Fe}_{6150}$, $A^{\rm Ba}_{6150}$, and $v_{\rm e}\sin i_{6150}$
(for the 6150 region); 
$A^{\rm C}_{5380}$, $A^{\rm Ti}_{5380}$, $A^{\rm Fe}_{5380}$, 
and $v_{\rm e}\sin i_{5380}$ (for the 5380 region); and
$A^{\rm O}_{7775}$, $A^{\rm Fe}_{7775}$, and $v_{\rm e}\sin i_{7775}$
(for the 7775 region).
In case that abundance solutions for some elements did not 
converge (especially for rapid rotators), we had to abandon
their determinations and fix them at the solar abundances 
(i.e., abundances used in the model atmosphere) during the 
iteration procedure and concentrate on the remaining parameters.
Figures 3 (6150 region), 4 (5380 region), and 5 (7775 region) 
show how the theoretical synthetic spectra corresponding to 
the final solutions fit the observations.

In order to demonstrate the importance (or unimportance) of the choice
of microturbulence for each element, we show the abundance differences 
between the two cases of $\xi^{\rm fit}$ and $\xi^{\rm std}$
in Figures 6a (6150 region) and b (7775 region) by using the 
abundance results obtained as by-products from the 
$\xi^{\rm fit}$-determinination mentioned in the previous Sect. IV-b.
It can be seen from these figures that the $A$ values are
not very sensitive to the choice of $\xi$, except that only $A^{\rm Ba}$
is considerably $\xi$-dependent because the Ba~{\sc ii} line 
at 6142.9~$\rm\AA$ (on which $A^{\rm Ba}$ essentially relies) is 
strongly saturated (cf. Figure 3). In the remainder of this paper, 
we exclusively refer to the abundances derived by using $\xi^{\rm std}$
as the standard abundances to be discussed.

We also estimated the uncertainties in $A^{\rm X}$
by repeating the analysis while perturbing
the standard values of the atmospheric parameters 
($T_{\rm eff}^{\rm std}$, $\log g^{\rm std}$, $\xi^{\rm std}$) 
interchangeably by $\pm 300$~K , $\pm 0.3$~dex, and 
$\pm 0.3 \xi^{\rm std}$~km~s$^{-1}$.\footnote{
We consider that typical uncertainties in $T_{\rm eff}$ and 
$\log g$ determinations for A-type stars are roughly on 
the order of $\sim 300$~K and $\sim 0.3$~dex, respectively, 
which we inferred from dispersions in the literature values of
stellar atmospheric parameters (e.g., Cayrel de Strobel, Soubiran, 
\& Ralite 2001 or Sadakane \& Okyudo 1989). Besides, as mentioned in 
Sect. IV-b, the ambiguity in $\xi$ is estimated to be $\pm 30\%$.}
Then, the root-sum-square of the resulting abundance changes 
($\Delta_{T}$, $\Delta_{g}$, $\Delta_{\xi}$) may be regarded as 
the error involved in $A^{\rm X}$; i.e., $\Delta A^{\rm X} \equiv 
(\Delta_{T}^{2} + \Delta_{g}^{2} + \Delta_{\xi}^{2})^{1/2}$

In discussing abundance peculiarities, it is useful
to represent the results of elemental abundances in terms 
of the differences relative to the fiducial values. 
Unfortunately, the Sun is not suitable for this purpose
since its spectrum appearance is considerably different
from that of A-type stars. Accordingly, we adopted
Procyon (F5 IV--V) as the reference star of abundance 
standard, considering that it has parameters not very different 
from those of A stars and its chemical composition is known to
be nearly the same as that of the Sun (cf., e.g., Kato \& Sadakane
1982, Steffen 1985, or Edvardsson et al. 1993; see also Figure 3 
in Varenne \& Monier 1999). Regarding the spectra of Procyon, 
we used Takeda et al.'s (2005a) OAO spectrum database for 
the 6160 and 5380 regions, while Allende Prieto et al.'s (2004)
public-domain spectrum was invoked for the 7775 region.
Adopting Takeda et al.'s (2005b) results for the atmospheric 
parameters ($T_{\rm eff} = 6612$~K, $\log g = 4.00$, and 
$\xi = 2.0$~km~s$^{-1}$), we derived the elemental abundances of
Procyon,\footnote{The resulting abundances of Procyon 
(in the usual normalization of $A^{\rm H} = 12.00$) are as follows:
$A^{\rm O}_{6150} = 8.87$, $A^{\rm Si}_{6150} = 7.14$, 
$A^{\rm Ca}_{6150} = 6.19$, $A^{\rm Fe}_{6150} = 7.49$, 
and $A^{\rm Ba}_{6150} = 2.33$ (for the 6150 region); 
$A^{\rm C}_{5380} = 8.75$, $A^{\rm Ti}_{5380} = 5.15$, 
and $A^{\rm Fe}_{5380} = 7.55$ (for the 5380 region); and
$A^{\rm O}_{7775} = 8.90$ and $A^{\rm Fe}_{7775} = 7.40$
(for the 7775 region).} 
from which [X/H] values (star$-$Procyon differential abundances) 
were computed as [X/H]$_{\rm region}$ $\equiv$ 
$A^{\rm X}_{region}$(star) $-$ $A^{\rm X}_{region}$(Procyon),
(X is any of C, O, Si, Ca, Ti, Fe, Ba; and {\it region} is
any of 6150, 5380, and 7775). Such obtained results of [X/H] are
presented in Table 1. 
Comparisons of [Fe/H]$_{5380}$ vs. 
[Fe/H]$_{6150}$, [Fe/H]$_{7775}$ vs. [Fe/H]$_{6150}$, and
[O/H]$_{7775}$ vs. [O/H]$_{6150}$ are shown in Figures 7a, b, and c,
respectively. 
We can see from these figures that the discrepancies tend to be
larger for rapid rotators, which indicates the growing difficulty
in abundance determinations of broad-line stars. 

From now on, in case where two or three kinds of 
solutions are available from different wavelength regions, we will use 
the result from the 6150 region (which is wider and include more 
lines than the other two).

The resulting rotational velocities ($v_{\rm e}\sin i_{6150}$) 
are compared with the previously published results by
Abt \& Morrell (1995) and Royer et al. (2002a, b) in Figure 8,
where we can recognize that our $v_{\rm e}\sin i$ solutions
are in reasonable agreement with these literature values.

\section{DISCUSSION}

\subsection{Rotation--Abundance Connection}

Figures 9a--f display the resulting [X/H] vs. [Fe/H] correlations 
(for X = C, O, Si, Ca, Ti, and Ba), from which we can 
roughly divide these elements into three groups.\\
(i) Si and Ti: almost scaling in accordance with Fe.\\
(ii) C, O, and Ca: showing an anti-correlation trend with Fe.\\
(iii) Ba: positive correlation with Fe, though its range of 
peculiarity is much more conspicuous than that of Fe.

Besides, [C/H], [O/H], [Si/H], [Ca/H], [Ti/H], [Ba/H], 
and [Fe/H] are plotted against $v_{\rm e} \sin i$ in Figures 10a--g.
We can recognize from these figures that [C/H], [O/H], [Ca/H] 
and [Ba/H] are systematically $v_{\rm e} \sin i$-dependent in the sense that
the peculiarity (overabundance for Ba, underabundance for C/O/Ca) 
tends to decrease with an increase in $v_{\rm e} \sin i$. 
While such a convincing tendency is not apparent for 
the remaining elements (Si, Ti, and Fe), [Fe/H] appears to 
weakly conform to this trend (i.e., decreasing tendency with 
$v_{\rm e} \sin i$).
 
Combining these observational fact, we may conclude as follows:\\
--- (a) All the seven elements exhibit some kind of abundance
peculiarities, which are more conspicuously seen in slow rotators 
($v_{\rm e}\sin i \la 50$~km~s$^{-1}$) and characterized by the deficiency 
of C, O, and Ca and the enrichment of Si, Fe, and (especially) Ba.\\
--- (b) These anomalies tend to diminish progressively with an
increase in $v_{\rm e} \sin i$ (at least in the range of slow/moderate
rotators of $\la 100$~km~s$^{-1}$). \\
--- (c) The stellar rotational velocity must thus be the most 
important key factor in the sense that the extent of abundance 
peculiarity tends to be larger as a star rotates more slowly,
which is presumably because some counter-acting mechanism of
diluting the built-up anomaly (most probably due to the
element segregation in a stable atmosphere/envelope) 
takes place in rapid rotators.

We also point out these tendencies seen in Figures 9 and 10 
are more or less consistent with the results of recently published 
papers focused on the abundance trends of A-type stars 
(including Am stars) for a wide range of $v_{\rm e} \sin i$ values: e.g.,
Lemke (1990, 1993) [field stars;  C, Ba (elements in common 
with this study)],
Savanov (1995a,b) [field stars; C, O, Si, Ca, Fe, Ba],  
Takeda \& Sadakane (1997) [Hyades and field stars; Fe, O],
Gebran, Monier, \& Richard (2008) [Coma Berenices; C, O, Si, Ca, Fe, Ba], 
Gebran \& Monier (2008) [Pleiades; C, O, Si, Ca, Fe, Ba],
and Fossati et al. (2008)[Praesepe; C, O, Si, Ca, Fe, Ba].

\subsection{Implication of Subsolar Compositions in Normal A Stars}

According to what we learned in Sect. V-a, we may assume
that the abundance peculiarities of A-type stars (conspicuously
seen slow rotators) tend to disappear for rapid rotators
at $v_{\rm e} \sin i \ga 100$~km~s$^{-1}$ (cf. Figure 10).
If so, we would be able to gain information of the galactic gas 
$\la 10^{9}$~yr ago by inspecting the photospheric abundances of 
such rapidly-rotating A-type stars, since they are considered to 
retain the composition of the gas from which they formed.

From this point of view, it is interesting to note in Figure 10 
that the [X/H] values at the high-$v_{\rm e} \sin i$ range tend to be 
somewhat negative or ``subsolar'' for many elements such as 
C, O, Ti, Fe, and Ba; i.e., by several tenths dex below the 
solar (or Procyon) abundances on the average.

Here we recall Takeda, Sato, \& Murata's (2008) conclusion that the 
[Fe/H] values (as well as those of other elements whose abundances 
almost scale with Fe) of evolved G giants, many of which have 
mass values around $\sim 2 M_{\odot}$ like A-type dwarfs, 
spread in a range of $-0.8 \la$~[Fe/H]~$\la +0.2$ around an average 
value of [Fe/H]~$\sim -0.3$.

Considering these two observational consequences, we would conclude
that the metallicities of the galactic gas $\la 10^{9}$~yr ago 
had really a subsolar tendency (though with a rather large diversity). 
If this is the case, the gas metallicity of [Fe/H] $\sim 0$ 
($\sim 5 \times 10^{9}$ ago when our Sun was born) must have decreased 
by several tenths dex with an elapse of time until $\la 10^{9}$~yr ago 
when A dwarfs (progenitors of G giants) were born. 
Although this trend does not seem to have been taken 
very seriously so far \footnote{Meanwhile, a completely different 
solution to this problem has also been proposed, arguing the necessity 
of downward revision of the solar abundances as a result of 
the application of sophisticated 3D line formation theory; 
(cf. Asplund et al. 2004). While this possibility may be 
worth considering, it can not yet be regarded as reliable 
in our opinion, since it causes serious discrepancies between 
theory and observation in the solar interior model (see, e.g., 
Young 2005 and the references therein). Besides, some questionable
points still remain in their line-formation treatment (see also 
Appendix 1 in Takeda \& Honda 2005).}
in spite of not a few supportive evidences\footnote{
Actually, the apparent subsolar tendency in the photospheric 
abundances of comparatively young stars has often been reported; 
e.g., C/N/O in early B main-sequence stars 
(Gies \& Lambert 1992, Kilian 1992, 
see also Nissen 1993); C/N/O/Si/Mg/Al in early B stars (Kilian 1994);
[Fe/H] in superficial normal late B and A stars (Sadakane 1990);  
[Fe/H] of B stars from UV spectra (Niemczura 2003);
O in supergiants (Luck \& Lambert 1985; Takeda \& Takada-Hidai 1998).}
since it contradicts the conventional scenario of galactic chemical
evolution (where elemental abundances are generally believed to increase 
with time), we tend to regard this tendency as real, which means that 
the gas metallicity actually {\it decreased} in an elapse of time
between the formation of our Sun ($\sim 5\times 10^{9}$~yr ago)
and the formation of $\sim 2~M_{\odot}$ stars ($\la 10^{9}$~yr ago).
Of course, in order to make this hypothesis more convincing, 
a reasonable explanation has to be done why such a reduction of 
the gas metallicity had occurred against the intuitive chemical evolution 
picture of increasing metallicity. While one such interpretation might be
the dilution of the metallicity caused by an substantial infall of 
metal-poor primordial galactic gas speculated by Takeda et al. (2008),
further observations and extensive abundance analyses on a much 
larger number of rapidly-rotating A dwarfs (as well as evolved 
G giants) would be required until we can say something about it 
with confidence.

\acknowledgments

We express our heartful thanks to Mr. Jin-Guk Seo for his technical 
support during the observations.

I. Han acknowledges the financial support for this study by KICOS through 
Korea-Ukraine joint research grant (grant  07-179).

B.-C. Lee acknowledges the Astrophysical Research Center for the Structure 
and Evolution of the Cosmos  (ARSEC, Sejong University) of the 
Korea Science and Engineering Foundation (KOSEF) through the Science
Research Center (SRC) program.


\newpage

\onecolumn

\setcounter{table}{0}
\begin{table}[h]
\caption{Basic stellar data and the resulting parameters and abundances.}
\scriptsize
\begin{center}
\begin{tabular}{cccccc@{  }c@{ }c@{ }c@{ }c@{ }c c@{ }c@{ }c c@{ }c}\hline\hline
HD & Sp. & $T_{\rm eff}$ & $\log g$ & $\xi^{\rm std}$ & $v_{\rm e} \sin i$ &
O & Si & Ca & Fe & Ba & C & Ti & Fe & O & Fe \\
   &     &  &  &  &  & 
\multicolumn{5}{c}{6150 region} & \multicolumn{3}{c}{5380 region} &
\multicolumn{2}{c}{7775 region} \\
\hline
130109 & A0V      &  9683 &  3.68 &  2.4 &  290 & $-$0.45 & +0.66 & $\cdots$ & $-$0.82 & $\cdots$   & $\cdots$ & $\cdots$ & $\cdots$   & $-$0.06 & $\cdots$\\
028024 & A8Vn     &  7107 &  3.20 &  3.3 &  250 & +0.13 & +0.05 & +0.04 & +0.03 & $-$0.29   & $\cdots$ & $-$0.21 & $-$0.71   & +0.15 & $\cdots$\\
106591 & A3V      &  8629 &  3.85 &  3.7 &  221 & $-$0.36 & $\cdots$ & $-$0.19 & $-$0.51 & $\cdots$   & +0.04 & $-$1.18 & $-$0.15   & $-$0.48 & $\cdots$\\
141003 & A3V      &  8580 &  3.56 &  3.7 &  220 & $-$0.10 & $\cdots$ & +0.80 & +0.03 & $\cdots$   & +0.12 & $-$0.51 & $-$0.14   & $-$0.31 & $\cdots$\\
027946 & A7V      &  7401 &  3.84 &  3.7 &  193 & +0.12 & +0.20 & +0.24 & +0.10 & $-$0.25   & $-$0.47 & +0.22 & $-$0.43   & +0.16 & $\cdots$\\
097603 & A4V      &  8180 &  3.90 &  4.0 &  191 & $-$0.01 & $\cdots$ & +0.44 & +0.00 & $-$0.51   & $-$0.07 & $-$0.37 & $-$0.30   & $-$0.28 & $\cdots$\\
059037 & A4V      &  8238 &  3.99 &  4.0 &  185 & +0.01 & $\cdots$ & +0.32 & +0.01 & $-$0.53   & $-$0.28 & $-$0.02 & $-$0.32   & $-$0.21 & $\cdots$\\
102124 & A4V      &  8026 &  4.09 &  4.0 &  185 & $-$0.06 & +0.37 & +0.40 & +0.08 & $-$0.22   & $-$0.26 & $-$0.23 & $-$0.40   & $-$0.25 & $-$0.23\\
080081 & A3V      &  9014 &  3.82 &  3.3 &  179 & $-$0.27 & $\cdots$ & +0.35 & $-$0.34 & $\cdots$   & $-$0.02 & $\cdots$ & $-$0.01   & $-$0.36 & $\cdots$\\
103287 & A0Ve     &  9202 &  3.79 &  3.0 &  164 & $-$0.18 & $\cdots$ & $-$0.36 & $-$0.26 & $\cdots$   & $-$0.38 & $\cdots$ & $-$0.25   & $-$0.40 & $\cdots$\\
056537 & A3V      &  8458 &  3.90 &  3.8 &  162 & +0.01 & $\cdots$ & +0.41 & $-$0.03 & $-$0.40   & $-$0.10 & $-$0.08 & $-$0.16   & $-$0.23 & +0.42\\
076644 & A7V      &  7934 &  4.22 &  4.0 &  139 & +0.03 & $-$0.30 & +0.06 & $-$0.21 & $-$0.50   & $-$0.26 & +0.03 & $-$0.30   & $-$0.27 & $-$0.43\\
029488 & A5Vn     &  7990 &  3.82 &  4.0 &  137 & +0.12 & $-$0.16 & $-$0.06 & $-$0.12 & $-$0.60   & $-$0.18 & $-$0.03 & $-$0.32   & $-$0.15 & $-$0.08\\
099211 & A7Vn:    &  7722 &  3.95 &  3.9 &  128 & $-$0.22 & $-$0.21 & $-$0.20 & $-$0.21 & $-$0.16   & $-$0.36 & $-$0.16 & $-$0.46   & $-$0.37 & $-$0.16\\
031295 & A0V      &  8993 &  4.11 &  3.3 &  123 & $-$0.33 & $\cdots$ & $-$0.70 & $-$0.56 & $\cdots$   & $-$0.29 & $\cdots$ & $-$0.71   & $-$0.20 & $\cdots$\\
127762 & A7III    &  7663 &  3.59 &  3.9 &  123 & $-$0.12 & $-$0.25 & $-$0.16 & $-$0.24 & +0.07   & $-$0.26 & $-$0.05 & $-$0.36   & $-$0.25 & $-$0.24\\
139006 & A0V      &  9573 &  3.87 &  2.5 &  121 & $-$0.17 & $\cdots$ & +0.12 & $-$0.15 & $-$0.80   & $-$0.07 & $-$0.76 & $-$0.13   & $-$0.54 & $\cdots$\\
028527 & A6IV     &  8039 &  3.99 &  4.0 &  120 & +0.15 & +0.00 & $-$0.08 & $-$0.06 & $-$0.54   & $-$0.13 & +0.11 & $-$0.26   & $-$0.03 & $-$0.36\\
032301 & A7V      &  7937 &  3.74 &  4.0 &  120 & +0.12 & $-$0.04 & $-$0.06 & $-$0.14 & $-$0.56   & $-$0.19 & +0.01 & $-$0.25   & $-$0.08 & $-$0.05\\
102647 & A3V      &  8643 &  4.17 &  3.7 &  120 & $-$0.06 & $-$0.46 & +0.01 & $-$0.12 & $-$0.16   & $-$0.24 & $-$0.20 & $-$0.11   & $-$0.28 & $\cdots$\\
028910 & A8V      &  7520 &  3.97 &  3.8 &  119 & $-$0.10 & $-$0.48 & $-$0.42 & $-$0.22 & $-$0.22   & $-$0.53 & $-$0.09 & $-$0.32   & $-$0.12 & +0.63\\
\hline
028355 & A7V      &  7809 &  3.98 &  4.0 &   91 & $-$0.07 & +0.11 & $-$0.41 & +0.12 & +0.42   & $-$0.50 & $-$0.08 & $-$0.19   & $-$0.27 & +0.00\\
027934 & A7IV-V   &  8159 &  3.84 &  4.0 &   87 & +0.07 & +0.14 & $-$0.10 & +0.01 & $-$0.37   & $-$0.13 & +0.04 & $-$0.22   & $-$0.12 & $-$0.07\\
074198 & A1IV     &  9381 &  4.11 &  2.8 &   87 & $-$0.31 & $-$0.04 & $-$0.15 & +0.26 & +0.63   & $-$0.39 & $-$0.11 & +0.24   & $-$0.25 & +0.26\\
033641 & A4m      &  7961 &  4.21 &  4.0 &   86 & $-$0.30 & $-$0.18 & $-$0.60 & +0.05 & +0.46   & $-$0.64 & +0.01 & $-$0.11   & $-$0.34 & $-$0.10\\
028226 & Am       &  7361 &  4.01 &  3.6 &   85 & $-$0.08 & +0.07 & $-$0.45 & +0.20 & +0.58   & $-$0.25 & $-$0.12 & $-$0.01   & $-$0.12 & +0.08\\
029388 & A6V      &  8194 &  3.88 &  4.0 &   85 & +0.02 & +0.08 & $-$0.08 & $-$0.05 & $-$0.36   & $-$0.17 & +0.03 & $-$0.25   & $-$0.10 & $-$0.03\\
025490 & A0.5Va   &  9077 &  3.93 &  3.2 &   80 & $-$0.50 & $\cdots$ & $-$0.29 & $-$0.10 & +0.11   & $-$0.50 & $-$0.45 & $-$0.12   & $-$0.29 & +0.70\\
028319 & A7III    &  7789 &  3.68 &  4.0 &   71 & $-$0.01 & +0.03 & $-$0.20 & $-$0.14 & $-$0.41   & $-$0.23 & +0.02 & $-$0.41   & $-$0.25 & $-$0.18\\
095382 & A5III    &  8017 &  3.95 &  4.0 &   70 & $-$0.03 & $-$0.15 & $-$0.11 & $-$0.15 & $-$0.06   & $-$0.27 & $-$0.05 & $-$0.30   & $-$0.01 & +0.32\\
116656 & A2V      &  9317 &  4.10 &  2.9 &   62 & $-$0.48 & $-$0.13 & $-$0.14 & +0.26 & +0.73   & $-$0.96 & +0.20 & +0.10   & $\cdots$ & $\cdots$\\
130841 & A5m$^{*}$&  8079 &  3.96 &  4.0 &   60 & $-$0.64 & $-$1.29 & $-$1.07 & $-$0.39 & $-$0.24   & $-$1.59 & $-$0.35 & $-$0.45   & $-$0.87 & +0.15\\
029479 & A4m      &  8406 &  4.14 &  3.9 &   58 & $-$0.15 & +0.13 & $-$0.40 & +0.29 & +1.08   & $-$0.44 & +0.18 & +0.14   & $-$0.12 & +0.31\\
089021 & A2IV     &  8861 &  3.61 &  3.5 &   52 & $-$0.25 & $-$0.06 & $-$0.25 & +0.06 & +0.53   & $-$0.57 & $-$0.08 & +0.02   & $-$0.30 & +0.23\\
\hline
027819 & A7V      &  8047 &  3.95 &  4.0 &   47 & $-$0.09 & $-$0.05 & +0.06 & $-$0.08 & $-$0.13   & $-$0.12 & $-$0.03 & $-$0.21   & $-$0.11 & $-$0.01\\
043378 & A2Vs     &  9210 &  4.09 &  3.0 &   46 & $-$0.13 & $-$0.47 & $-$0.07 & $-$0.17 & $-$0.17   & $-$0.29 & $-$0.05 & $-$0.21   & $-$0.18 & +0.18\\
095418 & A1V      &  9489 &  3.85 &  2.7 &   46 & $-$0.35 & $-$0.12 & $-$0.18 & +0.22 & +0.88   & $-$0.66 & $-$0.01 & +0.17   & $-$0.28 & +0.78\\
084107 & A2IV     &  8665 &  4.31 &  3.7 &   38 & $-$0.21 & $-$0.26 & $-$0.37 & $-$0.01 & +0.69   & $-$0.36 & +0.08 & $-$0.13   & $-$0.21 & $-$0.03\\
141795 & A2m      &  8367 &  4.24 &  3.9 &   34 & $-$0.68 & $-$0.09 & $-$0.67 & +0.18 & +1.15   & $-$0.98 & $-$0.15 & +0.07   & $-$0.60 & +0.16\\
028546 & Am       &  7640 &  4.17 &  3.9 &   28 & $-$0.30 & $-$0.03 & $-$0.40 & +0.12 & +0.83   & $-$0.46 & $-$0.07 & $-$0.07   & $-$0.33 & +0.11\\
095608 & A1m      &  8972 &  4.20 &  3.3 &   18 & $-$0.65 & $-$0.05 & $-$0.86 & +0.27 & +0.97   & $-$1.12 & $-$0.01 & +0.16   & $-$0.58 & +0.37\\
048915 & A1V      &  9938 &  4.31 &  2.1 &   17 & $-$0.40 & $-$0.21 & $-$0.46 & +0.40 & +1.20   & $-$0.91 & +0.15 & +0.41   & $-$0.31 & +0.49\\
072037 & A2m      &  7918 &  4.16 &  4.0 &   12 & $-$0.86 & +0.00 & $-$1.05 & +0.12 & +0.69   & $-$1.47 & $-$0.11 & $-$0.03   & $-$0.95 & +0.13\\
027962 & A2IV     &  8923 &  3.94 &  3.4 &   11 & $-$0.27 & +0.03 & $-$0.24 & +0.23 & +0.73   & $-$0.48 & +0.10 & +0.15   & $-$0.32 & +0.41\\
047105 & A0IV     &  9115 &  3.49 &  3.2 &   11 & $-$0.05 & $-$0.12 & $-$0.03 & $-$0.05 & +0.19   & $-$0.28 & $-$0.02 & $-$0.07   & $-$0.41 & +0.26\\
040932 & A2V      &  8005 &  3.93 &  4.0 &   10 & $-$0.55 & $-$0.44 & $-$0.77 & $-$0.20 & +0.23   & $-$0.66 & $-$0.37 & $-$0.45   & $-$0.25 & $-$0.21\\
\hline
\end{tabular}
\end{center}
In columns 1 through 5 are given the HD number, spectral type (from SIMBAD database),
effective temperature (in K), logarithmic surface gravity (in  cm~s$^{-2}$), and
microturbulent velocity (in km~s$^{-1}$). Columns 6 through 11 show the
results determined from 6150 region fitting: the projected rotational velocity
(in km~s$^{-1}$), [O/H], [Si/H], [Ca/H], [Fe/H], and [Ba/H]. Similarly,
the abundance results from the 5380 region fitting ([C/H], [Ti/H], and [Fe/H])
and the 7775 region fitting ([O/H] and [Fe/H]) are given in columns 12--14
and 15--16, respectively. All abundance results ([X/H]) are the differential values 
relative to Procyon. The 46 stars are arranged in the descending order of
$v_{\rm e} \sin i$, which are divided into three groups: rapidly-rotating
stars ($v_{\rm e} \sin i > 100$~km~s$^{-1}$), moderately-rotating stars
(50~km~s$^{-1} < v_{\rm e} \sin i < 100$~km~s$^{-1}$), and slowly-rotating stars
($v_{\rm e} \sin i < 50$~km~s$^{-1}$). \\ \\
$^{*}$  More exactly, SIMBAD gives ``kA2hA5mA4Iv-v'' for the spectral type of this star.
\end{table}

\newpage

\setcounter{table}{1}
\begin{table}[h]
\caption{Atomic data of important lines.$^{*}$}
\small
\begin{center}
\begin{tabular}{ccrc}\hline\hline
Species & $\lambda$ ($\rm\AA$) & $\chi$ (eV) & $\log gf$ \\
\hline
Ba~{\sc ii} & 6142.928 &  0.552 & $-$0.992 \\
Si~{\sc i} & 6143.125 &  5.964 & $-$2.790 \\
Si~{\sc i} & 6145.016 &  5.616 & $-$0.820 \\
Fe~{\sc ii} & 6147.741 &  3.889 & $-$2.721 \\
Fe~{\sc ii} & 6149.258 &  3.889 & $-$2.724 \\
Fe~{\sc i} & 6151.617 &  2.176 & $-$3.299 \\
Si~{\sc i} & 6155.134 &  5.619 & $-$0.400 \\
O~{\sc i} & 6155.961 & 10.740 & $-$1.401 \\
O~{\sc i} & 6155.971 & 10.740 & $-$1.051 \\
O~{\sc i} & 6155.989 & 10.740 & $-$1.161 \\
O~{\sc i} & 6156.737 & 10.740 & $-$1.521 \\
O~{\sc i} & 6156.755 & 10.740 & $-$0.931 \\
O~{\sc i} & 6156.778 & 10.740 & $-$0.731 \\
O~{\sc i} & 6158.149 & 10.741 & $-$1.891 \\
O~{\sc i} & 6158.172 & 10.741 & $-$1.031 \\
O~{\sc i} & 6158.187 & 10.741 & $-$0.441 \\
Ca~{\sc i} & 6161.297 &  2.523 & $-$1.020 \\
Ca~{\sc i} & 6162.173 &  1.899 &  +0.100 \\
Ca~{\sc i} & 6163.755 &  2.521 & $-$1.020 \\
Fe~{\sc i} & 6165.361 &  4.143 & $-$1.550 \\
Ca~{\sc i} & 6166.439 &  2.521 & $-$0.900 \\
\hline
Fe~{\sc i} & 5379.574 &  3.695 & $-$1.480 \\
C~{\sc i} & 5380.224 &  8.850 & $-$2.030 \\
C~{\sc i} & 5380.265 &  8.850 & $-$2.820 \\
C~{\sc i} & 5380.265 &  8.850 & $-$2.820 \\
Ti~{\sc ii} & 5381.015 &  1.566 & $-$2.080 \\
Fe~{\sc i} & 5383.369 &  4.312 &  +0.500 \\
Fe~{\sc i} & 5386.335 &  4.154 & $-$1.770 \\
Fe~{\sc i} & 5386.959 &  3.642 & $-$2.624 \\
Fe~{\sc ii} & 5387.063 & 10.521 &  +0.518 \\
Fe~{\sc i} & 5387.488 &  4.143 & $-$2.140 \\
\hline
O~{\sc i} & 7771.944 &  9.146 &  +0.324 \\
O~{\sc i} & 7774.166 &  9.146 &  +0.174 \\
O~{\sc i} & 7775.388 &  9.146 & $-$0.046 \\
Fe~{\sc ii} & 7780.354 &  9.761 & $-$0.547 \\
Fe~{\sc i} & 7780.552 &  4.473 & $-$0.066 \\
\hline
\end{tabular}
\end{center}
$^{*}$ All data were taken from the compilation of Kurucz \& Bell (1995),
except for the $gf$ value of the Fe~{\sc i} line at 7780.552~$\rm\AA$,
for which we used Kurucz \& Peytremann's (1975) value 
in accordance with Takeda \& Sadakane (1997).
\end{table}

\newpage

\setcounter{figure}{0}
\begin{figure}
  \begin{center}
    \includegraphics[width=8.0cm]{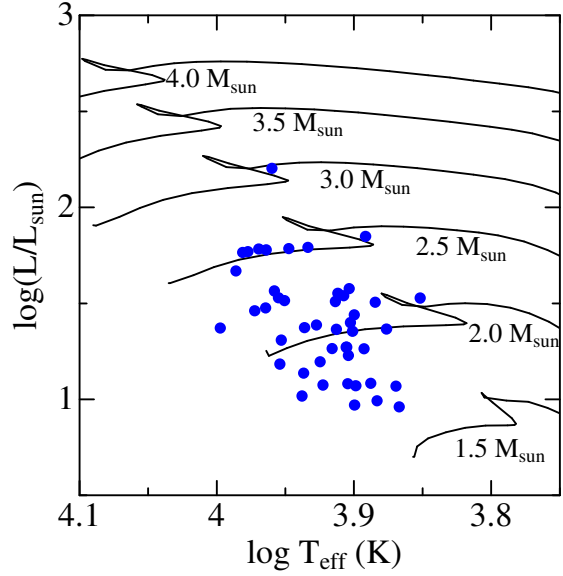}
  \end{center}
\caption{Plots of 46 program stars on the theoretical HR diagram
($\log (L/L_{\odot})$ vs. $\log T_{\rm eff}$ ), where the bolometric 
luminosity ($L$) was evaluated from the apparent visual magnitude with 
the help of Hipparcos parallax (ESA 1997) and Flower's (1996) bolometric
correction. Theoretical evolutionary tracks corresponding to 
the solar metallicity computed by Girardi et al. (2000) for 
six different initial masses are also depicted for comparison.}
\label{fig1}
\end{figure}

\setcounter{figure}{1}
\begin{figure}
  \begin{center}
    \includegraphics[width=15.0cm]{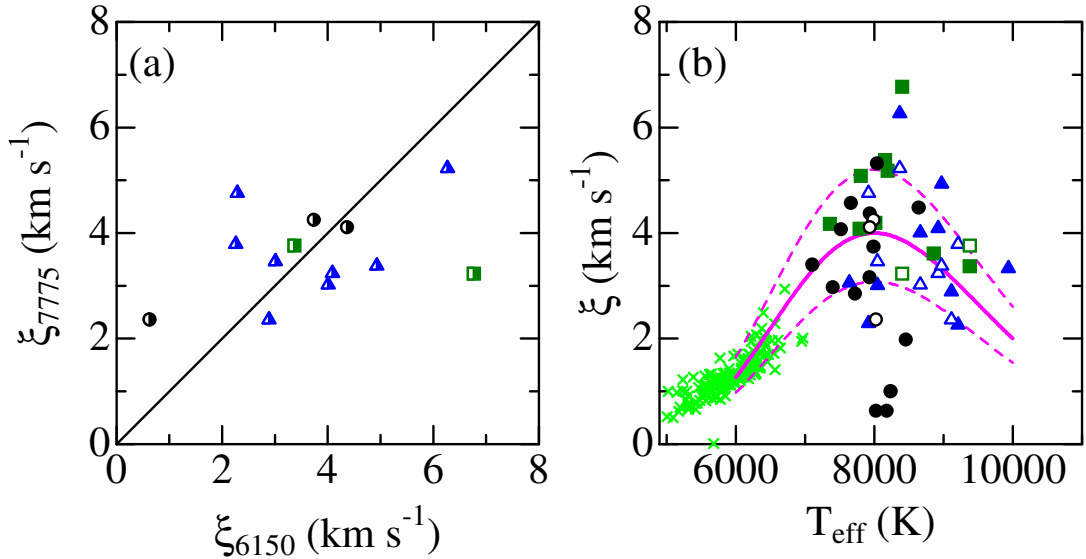}
  \end{center}
\caption{(a) Comparison of the microturbulence
derived from the 7775 region ($\xi_{7775}$) with that from
the 6150 region ($\xi_{6150}$).
(b) Microturbulences plotted 
against the effective temperature, where $\xi_{6150}$ and 
$\xi_{7775}$ are denoted by filled and open symbols, respectively.
The results for F--G--K dwarfs taken from Takeda et al. (2005b) 
are also shown by crosses for comparison.
In both panels, three groups of projected rotational velocity are
discerned by the symbol shape: Triangles $\cdots$ slow rotators
(0~km~s$^{-1} < v_{\rm e} \sin i <$ 50~km~s$^{-1}$), 
squares $\cdots$ moderate rotators
(50~km~s$^{-1} < v_{\rm e} \sin i <$ 100~km~s$^{-1}$), and 
circles $\cdots$ rapid rotators
(100~km~s$^{-1} < v_{\rm e} \sin i$).
The adopted $\xi$ vs. $T_{\rm eff}$ relation [cf. Equation (1)] is 
depicted by solid line, while its reasonable upper and lower limits 
(perturbations by $\pm 30\%$ as possible range of errors) are 
shown by two dashed lines. 
}
\label{fig2}
\end{figure}

\setcounter{figure}{2}
\begin{figure}
  \begin{center}
    \includegraphics[width=15.0cm]{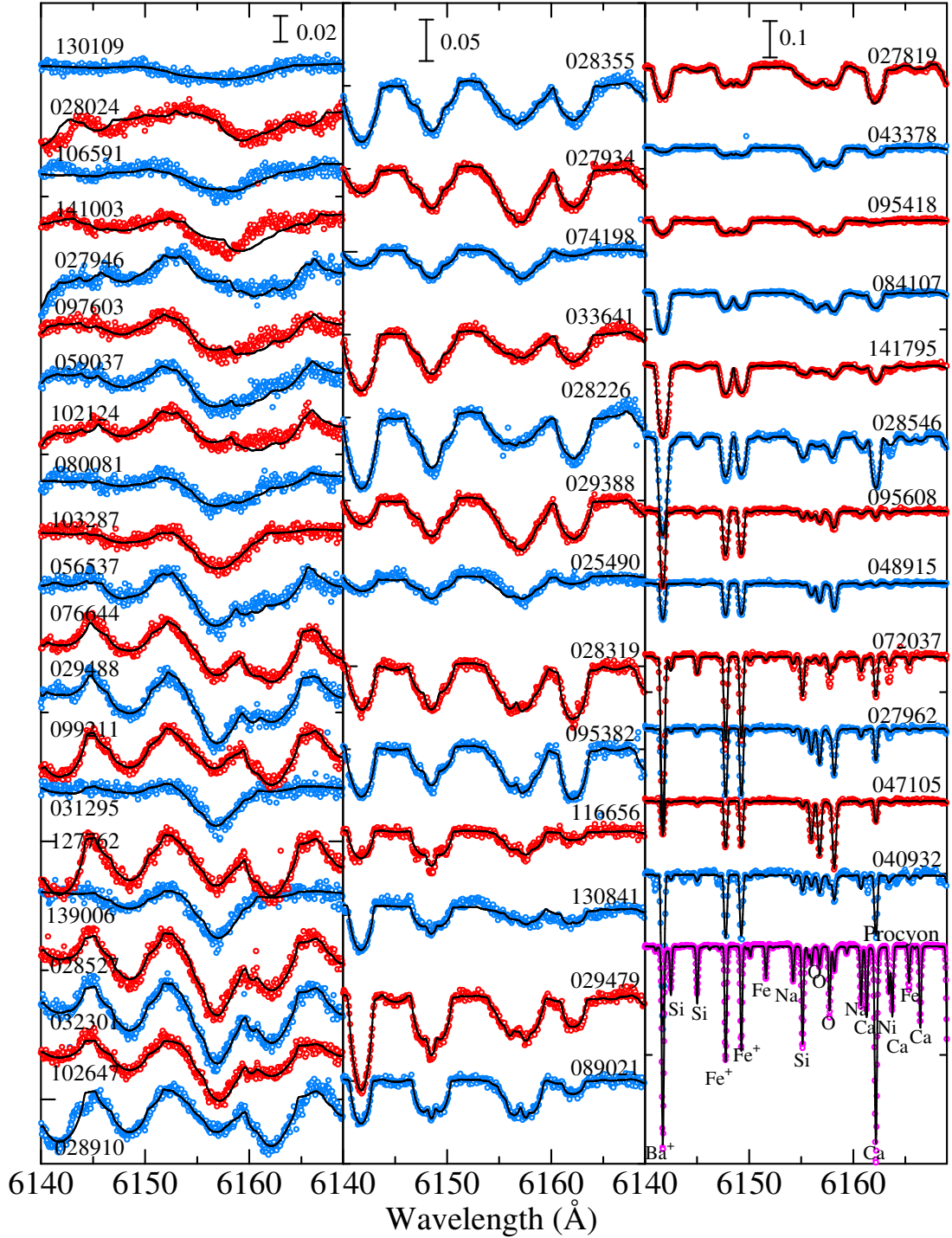}
  \end{center}
\caption{Synthetic spectrum fitting at the 6150 region 
(6140--6170~$\rm\AA$) for determining the projected rotational 
velocity ($v_{\rm e} \sin i$) and the abundances of O, Si, Ca, Fe, and Ba.
The best-fit theoretical spectra are shown by solid lines, 
while the observed data are plotted by symbols.  
Left panel $\cdots$ rapid rotators
(100~km~s$^{-1} < v_{\rm e} \sin i$), middle panel $\cdots$ 
moderate rotators (50~km~s$^{-1} < v_{\rm e} \sin i <$ 100~km~s$^{-1}$),
and right panel $\cdots$ slow rotators
(0~km~s$^{-1} < v_{\rm e} \sin i <$ 50~km~s$^{-1}$). In each panel,
the spectra are arranged in the descending order of $v_{\rm e} \sin i$,
and an appropriate offset is applied to each spectrum (indicated
by the HD number) relative to the adjacent one. The case of Procyon 
(standard star) is displayed at the bottom of the right panel.
}
\label{fig3}
\end{figure}

\setcounter{figure}{3}
\begin{figure}
  \begin{center}
    \includegraphics[width=15.0cm]{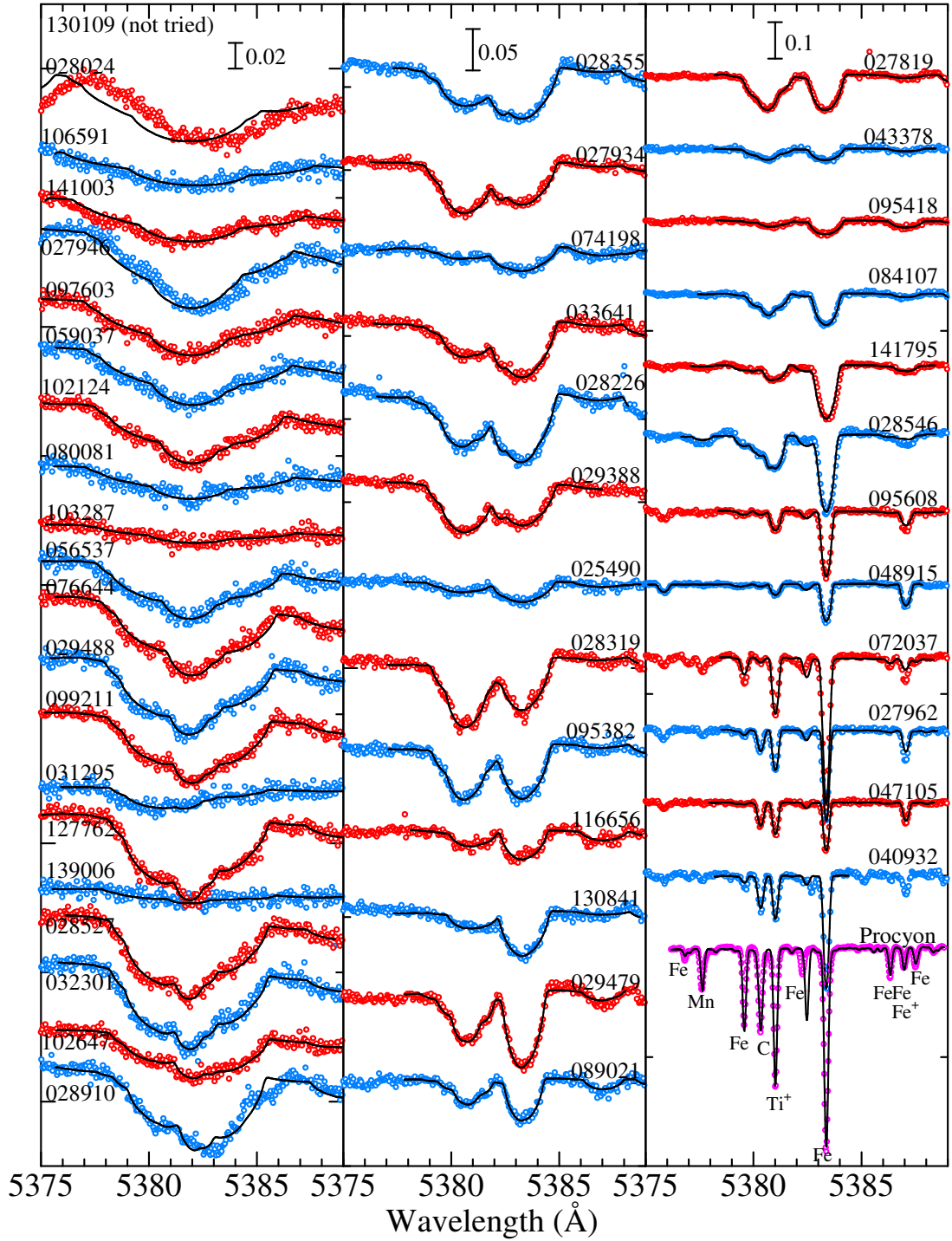}
  \end{center}
\caption{Synthetic spectrum fitting at the 5380 region 
(5375--5390~$\rm\AA$) for determining $v_{\rm e} \sin i$ and the 
abundances of C, Ti, and Fe. Otherwise, the same as in Fig. 3.}
\label{fig4}
\end{figure}

\setcounter{figure}{4}
\begin{figure}
  \begin{center}
    \includegraphics[width=15.0cm]{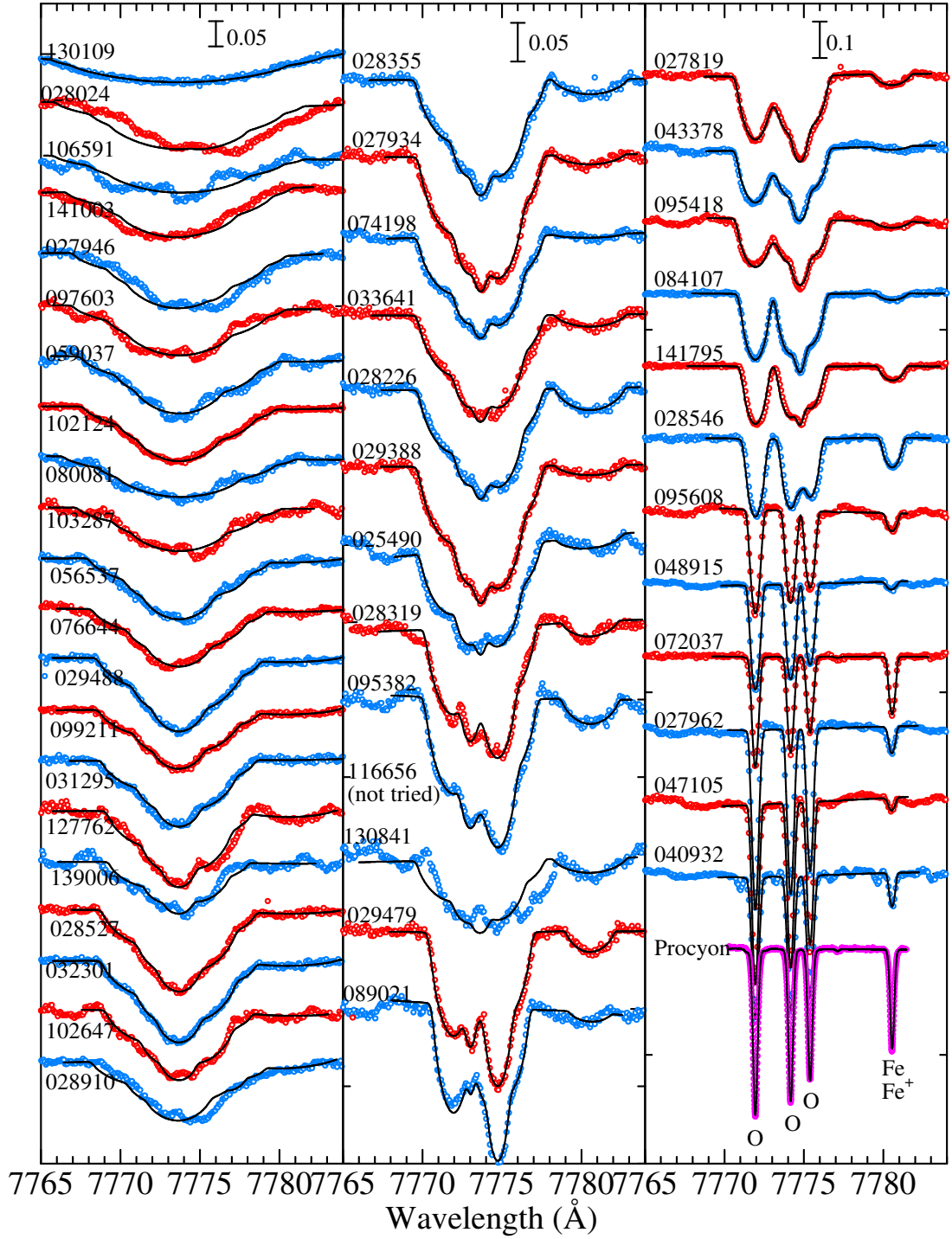}
  \end{center}
\caption{Synthetic spectrum fitting at the 7775 region 
(7765--7785~$\rm\AA$) for determining $v_{\rm e} \sin i$ and 
the abundances of O (with the non-LTE effect taken into account) 
and Fe. Otherwise, the same as in Fig. 3.}
\label{fig5}
\end{figure}

\setcounter{figure}{5}
\begin{figure}
  \begin{center}
    \includegraphics[width=15.0cm]{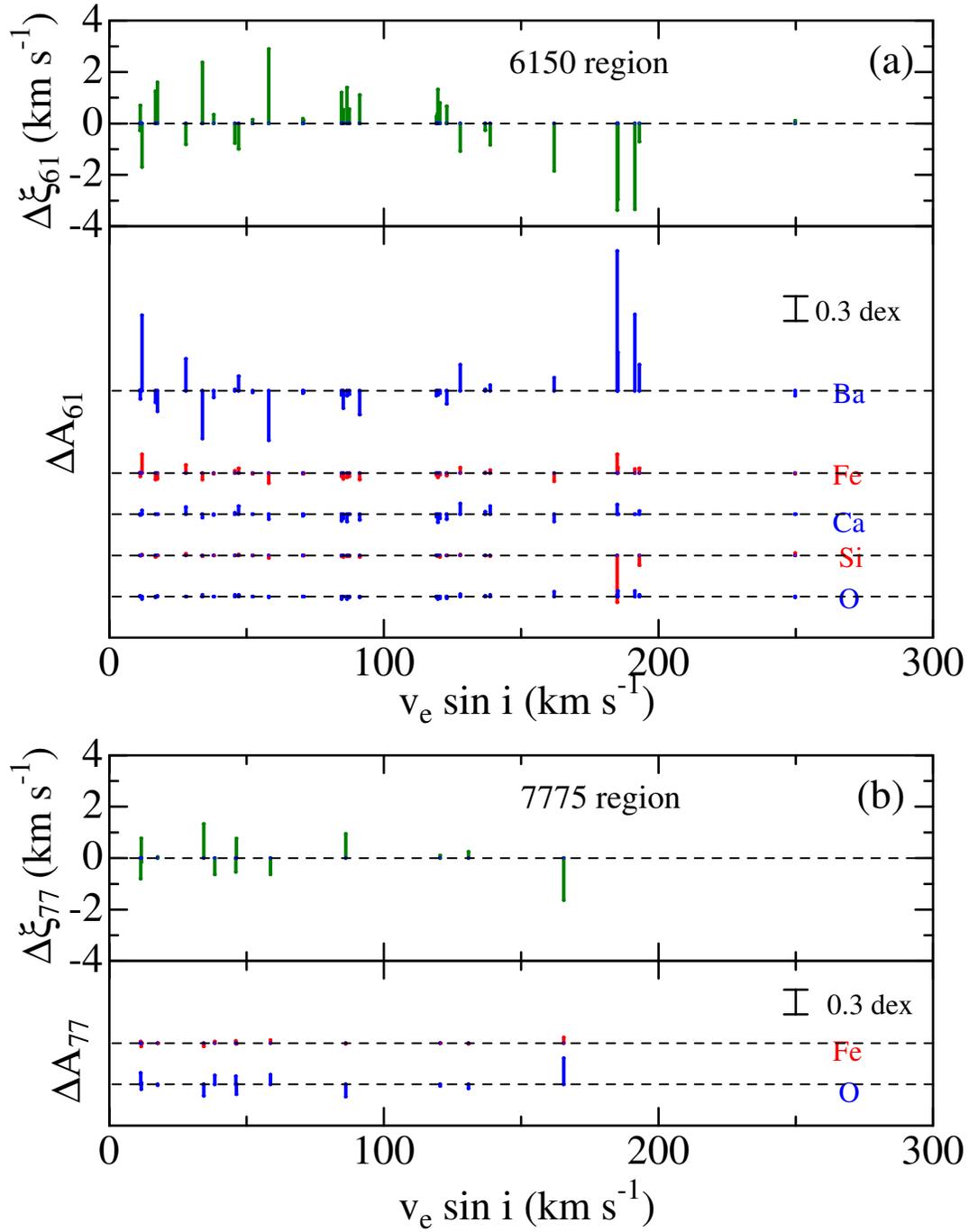}
  \end{center}
\caption{Abundance changes between the two cases of 
$\xi^{\rm fit}$ and $\xi^{\rm std}$ 
[$\Delta A \equiv A(\xi^{\rm fit}) - A(\xi^{\rm std})$]
plotted as functions of $v_{\rm e} \sin i$. (a) 6150 region
(33 stars) and (b) 7775 region (13 stars). 
In each figure, the upper and lower
panels show $\Delta \xi \equiv (\xi^{\rm fit} -\xi^{\rm std})$
and $\Delta A$ (for each element), respectively.}
\label{fig6}
\end{figure}

\setcounter{figure}{6}
\begin{figure}
  \begin{center}
    \includegraphics[width=15.0cm]{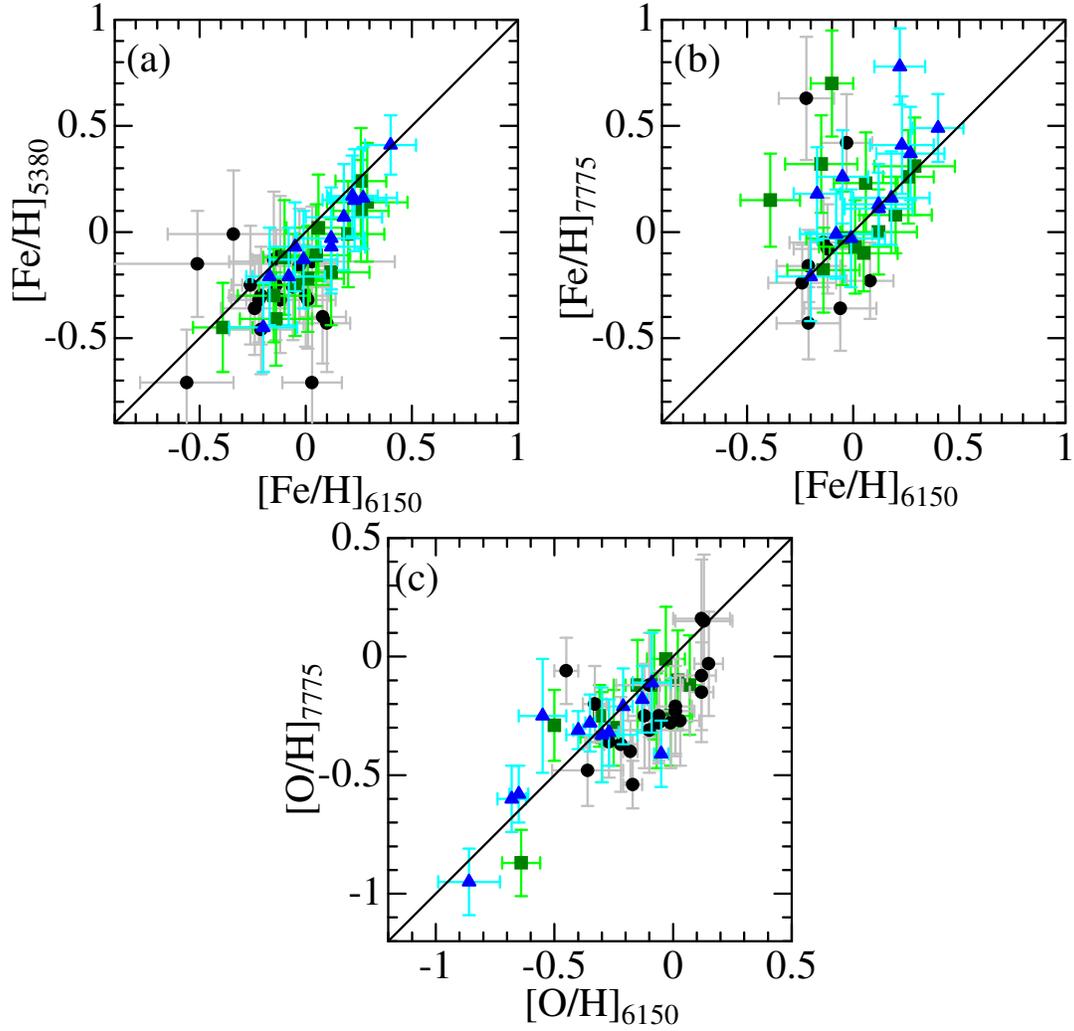}
  \end{center}
\caption{Comparisons of the abundances derived from different 
regions (cf. Table 1). (a) [Fe/H]$_{5380}$ vs. [Fe/H]$_{6150}$,
(b) [Fe/H]$_{7775}$ vs. [Fe/H]$_{6150}$, 
and (c) [O/H]$_{7775}$ vs. [O/H]$_{6150}$. 
The error bars attached in [X/H] represent the values of 
$\Delta A^{\rm X}$ (see Sect. IV-c).
Three groups of different $v_{\rm e} \sin i$ ranges are discriminated 
by the symbol shape in the same manner as in Fig. 2.}
\label{fig7}
\end{figure}

\setcounter{figure}{7}
\begin{figure}
  \begin{center}
    \includegraphics[width=15.0cm]{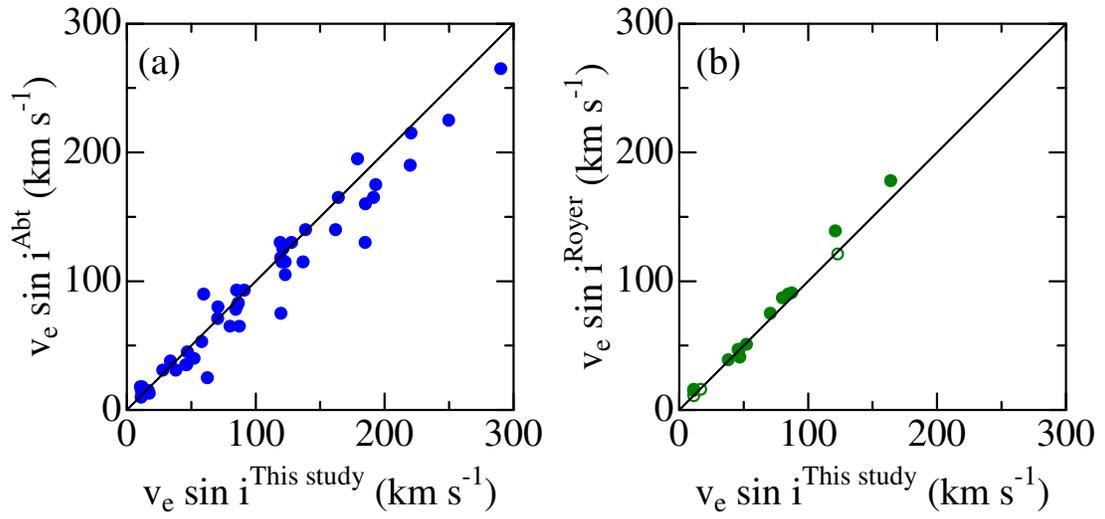}
  \end{center}
\caption{Comparison of the $v_{\rm e} \sin i$ results derived in this study 
(determined from the fitting in the 6150 region; cf. Table 1).
with the literature values: (a) Abt \& Morrell (1995), (b)
Royer et al. (2002a; open symbols; southern hemisphere) and 
Royer et al. (2002b; filled symbols; northern hemisphere).}
\label{fig8}
\end{figure}

\setcounter{figure}{8}
\begin{figure}
  \begin{center}
    \includegraphics[width=15.0cm]{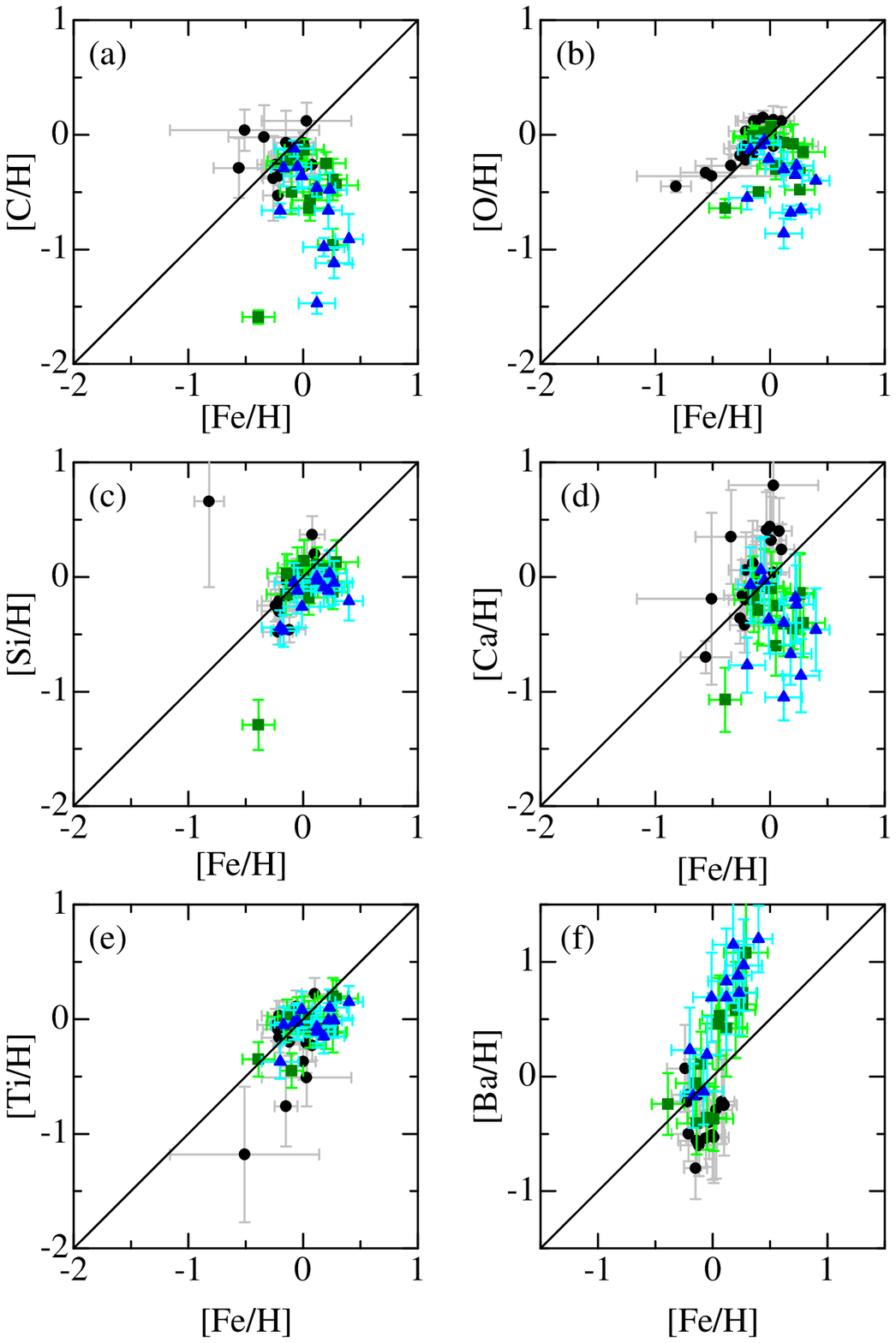}
  \end{center}
\caption{[X/H] values plotted against [Fe/H] (6150 region). 
(a) [C/H], (b) [O/H] (6150 region), (c) [Si/H], (d) [Ca/H], 
(e) [Ti/H], and (f) [Ba/H]. 
The error bars attached in [X/H] represent the values of 
$\Delta A^{\rm X}$ (see Sect. IV-c).
See the caption of Fig. 2 for the meanings of the symbols.}
\label{fig9}
\end{figure}

\setcounter{figure}{9}
\begin{figure}
  \begin{center}
    \includegraphics[width=15.0cm]{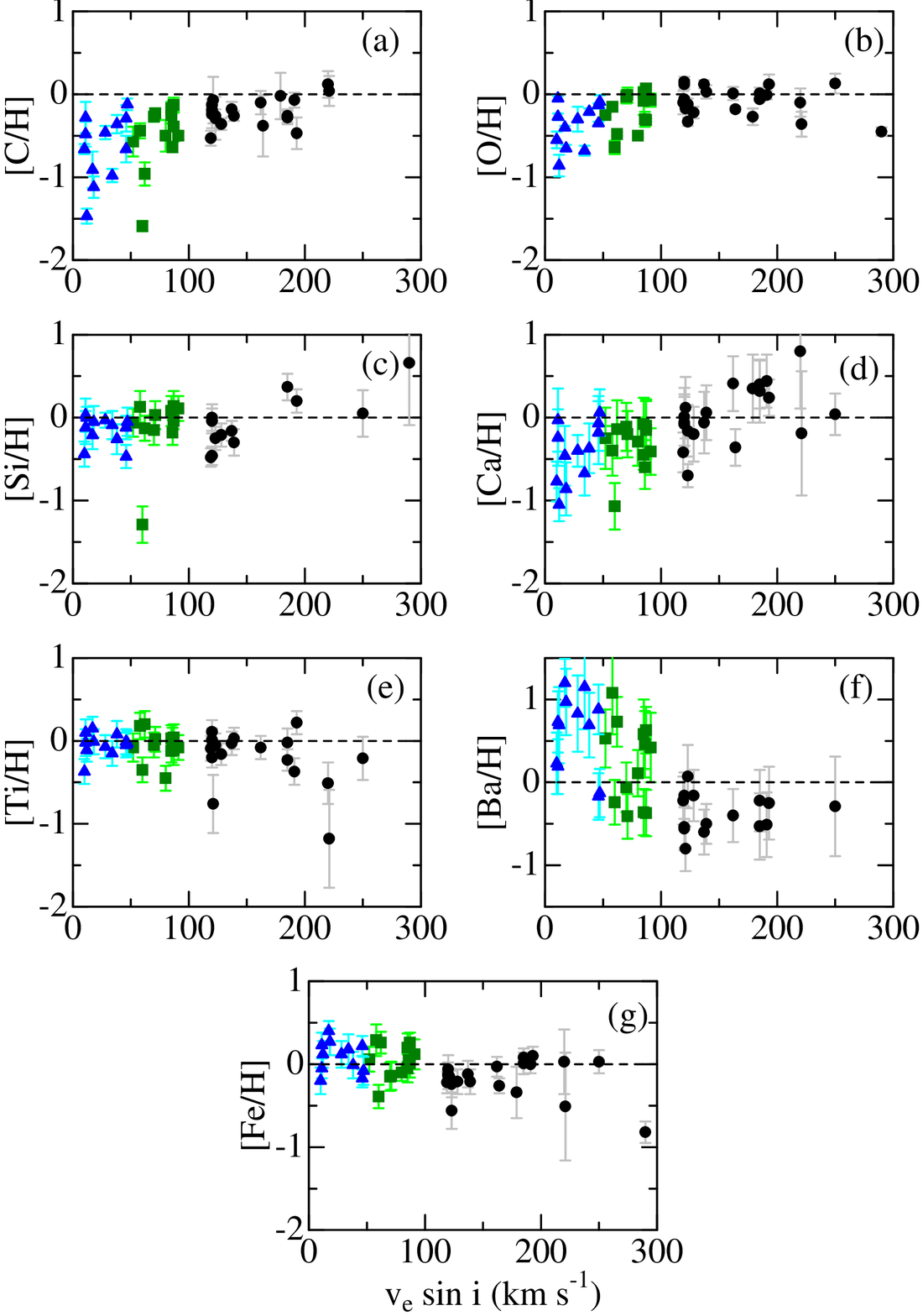}
\caption{[X/H] values plotted against $v_{\rm e} \sin i$ (6150 region).
(a) [C/H], (b) [O/H] (6150 region), (c) [Si/H], (d) [Ca/H], 
(e) [Ti/H], (f) [Ba/H], and (g) [Fe/H] (6150 region). 
The error bars attached in [X/H] represent the values of 
$\Delta A^{\rm X}$ (see Sect. IV-c).
See the caption of Fig. 2 for the meanings of the symbols.}
  \end{center}
\label{fig10}
\end{figure}

\end{document}